\documentclass{webofc}
\usepackage[varg]{txfonts}   
\usepackage[breaklinks, citecolor=blue, linkcolor=blue, colorlinks=true, debug, baseurl=' ']{hyperref}
\usepackage{graphicx, color}
\usepackage{xcolor}

\begin{document}

\title{CONCERTO: instrument and status}

\author{
\lastname{Alessandro Fasano}\inst{\ref{lam},\ref{iac},\ref{ull}}\fnsep\thanks{\email{alessandro.fasano@iac.es}}
\and
\lastname{Peter~Ade} \inst{\ref{cardiff}}
\and 
\lastname{Manuel~Aravena} \inst{\ref{santiago}}
\and 
\lastname{Emilio~Barria} \inst{ \ref{neel} }
\and 
\lastname{Alexandre~Beelen} \inst{\ref{lam}}
\and 
\lastname{Alain~Beno\^it} \inst{\ref{neel}}
\and 
\lastname{Matthieu~B\'ethermin} \inst{\ref{lam}}
\and 
\lastname{Julien~Bounmy} \inst{ \ref{lpsc} }
\and 
\lastname{Olivier~Bourrion} \inst{\ref{lpsc}}
\and 
\lastname{Guillaume~Bres} \inst{\ref{neel}}
\and 
\lastname{Martino~Calvo} \inst{\ref{neel}}
\and
\lastname{Andrea~Catalano} \inst{\ref{lpsc}}
\and 
\lastname{Carlos~De Breuck} \inst{\ref{eso_de}}
\and 
\lastname{François-Xavier~D\'esert} \inst{\ref{ipag}}
\and 
\lastname{Carlos~Dur\'an} \inst{\ref{eso_ch}}
\and 
\lastname{Thomas~Fenouillet} \inst{\ref{lam}}
\and 
\lastname{Jose~Garcia} \inst{\ref{lam}}
\and 
\lastname{Gregory~Garde} \inst{\ref{neel}}
\and 
\lastname{Johannes~Goupy} \inst{\ref{neel}}
\and 
\lastname{Christopher~Groppi} \inst{\ref{arizona}}
\and 
\lastname{Christophe~Hoarau} \inst{ \ref{lpsc} }
\and 
\lastname{Wenkai~Hu} \inst{\ref{lam}}
\and 
\lastname{Guilaine~Lagache} \inst{\ref{lam}}
\and 
\lastname{Jean-Charles~Lambert} \inst{\ref{lam}}
\and 
\lastname{Jean-Paul~Leggeri} \inst{\ref{neel}}
\and 
\lastname{Florence~Levy-Bertrand} \inst{\ref{neel}}
\and
\lastname{Andreas~Lundgren} \inst{\ref{lam}}
\and 
\lastname{Juan~Mac\'{\i}as-P\'erez} \inst{ \ref{lpsc} }
\and
\lastname{Hamdi~Mani} \inst{\ref{arizona}}
\and 
\lastname{Julien~Marpaud} \inst{\ref{lpsc}}
\and 
\lastname{Philip~Mauskopf} \inst{\ref{arizona}}
\and 
\lastname{Alessandro~Monfardini} \inst{\ref{neel}}
\and 
\lastname{Giampaolo~Pisano} \inst{\ref{cardiff}}
\and 
\lastname{Nicolas~Ponthieu} \inst{ \ref{ipag} }
\and 
\lastname{Leo~Prieur} \inst{\ref{lam}}
\and 
\lastname{Samuel~Roni} \inst{\ref{lpsc}}
\and 
\lastname{Sebastien~Roudier} \inst{\ref{lpsc}}
\and 
\lastname{Damien~Tourres} \inst{\ref{lpsc}}
\and 
\lastname{Carol~Tucker} \inst{\ref{cardiff}} 
}

\institute{
Aix Marseille Univ., CNRS, CNES, LAM, Marseille, France \label{lam}
\and 
Instituto de Astrof\'sica de Canarias, E-38205 La Laguna, Tenerife, Spain \label{iac}
\and
Departamento de Astrofísica, Universidad de La Laguna (ULL), E-38206 La Laguna, Tenerife, Spain \label{ull}
\and
Astronomy Instrumentation Group, University of Cardiff, The Parade, CF24 3AA, United Kingdom \label{cardiff}
\and
N\'ucleo de Astronom\'ia, Facultad de Ingenier\'ia y Ciencias, Universidad Diego Portales, Av.  Ej\'ercito 441, Santiago, Chile \label{santiago}
\and
Univ. Grenoble Alpes, CNRS, Grenoble INP, Institut N\'eel, 38000 Grenoble, France
\label{neel}
\and
Univ. Grenoble Alpes, CNRS, LPSC/IN2P3, 38000 Grenoble, France \label{lpsc}
\and
European Southern Observatory, Karl Schwarzschild Straße 2, 85748 Garching, Germany \label{eso_de}
\and
European Southern Observatory, Alonso de Cordova 3107, Vitacura, Santiago, Chile \label{eso_ch}
\and
Univ. Grenoble Alpes, CNRS, IPAG, 38400 Saint Martin d'H\'eres, France \label{ipag}
\and
School of Earth and Space Exploration and Department of Physics, Arizona State University, Tempe, AZ 85287, USA \label{arizona}
} 

\abstract{
CONCERTO (CarbON CII line in post-rEionization and ReionizaTiOn) is a low-resolution Fourier transform spectrometer dedicated to the study of star-forming galaxies and clusters of galaxies in the transparent millimeter windows from the ground. It is characterized by a wide instantaneous 18.6\,arcmin field of view, operates at 130--310\,GHz, and was installed on the 12-meter Atacama Pathfinder Experiment (APEX) telescope at 5\,100\,m above sea level. CONCERTO's double focal planes host two arrays of 2\,152 kinetic inductance detectors and represent a pioneering instrument to meet a state-of-the-art scientific challenge. This paper introduces the CONCERTO instrument and explains its status, shows the first CONCERTO spectral maps of Orion, and describes the perspectives of the project.}

\maketitle

\section{Introduction}
\label{sec:intro}

Modern study of the reionization and post-reionization epochs of the Universe and galaxy-cluster characterization requires the design of ambitious instrumentation in the millimeter (mm) domain. Next-generation satellites dedicated to mm wavelengths are currently being studied for the coming decades. In the meantime, the international community is developing ground-based experiments to improve our knowledge of the topics mentioned and define new technologies that the satellite experiments would ultimately adopt. The study of the reionization and post-reionization epochs of the Universe can be pursued via the measurement of  [CII] line emission \cite{lagache}. [CII] is among the brightest lines originating from star-forming galaxies and is a reliable tracer of star formation on a global scale. It is red-shifted in the transparent mm atmospheric windows at $z>5.2$ \cite{kovetz,concerto} and observed via 3-D line intensity mapping. This technique adds the spectral information in the third dimension in addition to the two sky coordinates, and follows the line evolution with redshift. It measures the angular fluctuations in sky brightness rather than solving the single galaxy structure and analyzes the [CII] distribution on different spatial scales.
[CII] intensity mapping is also one of the main goals of CCAT-prime \cite{choi} and TIME \cite{crites}. 

Galaxy clusters are now commonly observed at mm wavelengths via the Sunyaev--Zel’dovich effect (SZ). The complementary mm information, together with traditional X-ray observations, permits a more detailed comprehension of cluster structure and its physical characterization, and adds cosmological interest to galaxy-cluster science. Determining the distribution of clusters of galaxies as a function of mass and redshift has proven to be a powerful cosmological tool, particularly for inferring the matter content and dynamics of the Universe. Again, a strategy for studying galaxy clusters requires a multi-frequency capability by aiming to fully separate the different components and extract physical information (pressure, temperature, density, and line-of-sight velocity).

The mapping of [CII] at high redshift and galaxy clusters via SZ can be achieved at same electromagnetic frequencies. Both scientific targets demand high mapping speed combined with a multi-frequency capability. CONCERTO was conceived to fulfill these needs; it requires a wide field-of-view (FoV) Fourier transform spectrometer (FTS) operating at mm wavelengths.

This paper presents the CONCERTO instrument in Sect.~\ref{sec:instrument}. In Sect.~\ref{sec:orion}, we present the first CONCERTO spectral maps of the Orion Nebula. Section~\ref{sec:perspectives} illustrates the project's timeline, status, and perspectives.

\section{The instrument and its main systematic effects}
\label{sec:instrument}

The CONCERTO instrument generates more than 16k spectra per second during observations by performing on-the-fly (otf) scanning at tens of arcseconds per second with an instantaneous FoV of 18.6\,arcmin (see \citealp{fasano_aa,fasano_nika2} for more details of the otf scanning technique while continuously recording with the FTS). First, we must maximize the mapping speed to guarantee the scientifically required coverage of a  few square degrees of sky demanded by the [CII] intensity mapping. Secondly, the sampling rate has to be fast enough to compensate for atmospheric fluctuations. CONCERTO comprises a fast FTS that records a full spectrum of the projected FoV in sub-second time intervals, where the knee point of 1/f atmospheric noise arises at $\lesssim 1$\,Hz. A linear motor creates the optical path difference (OPD) within the FTS by oscillating the position of a roof mirror, and its stroke length defines the spectral resolution. This design creates an interference figure called an interferogram. The single detector recovers all the spectral information by scanning the interferogram. In fact, rather than the single detector efficiency, the gain is obtained by adopting large format arrays, in other words, by increasing the mapping speed. The FTS permits the exploitation of large-array formats multiplexing the electromagnetic frequencies. For CONCERTO, 4\,304 kinetic inductance detectors (KIDs; \citealp{doyle}) are projected onto the sky and distributed in two focal planes. These are orthogonally positioned facing the two sides of the last FTS polarizer which differentiates the optical path in a transmitted and reflected direction corresponding to the two focal planes. Two arrays, the low-frequency (LF) and the high-frequency (HF) array, are installed at the focal plane positions. The two arrays share the same design but are distinguishable by two different filtering chains, resulting in overlapping bandpasses that do not cover the same electromagnetic regions.

The instrumental characteristics of CONCERTO measured during the 2021 commissioning campaign on the sky are summarized in Table~\ref{tab:specs}. 

\begin{table}[ht]
\caption{Main characteristics of CONCERTO instrument coupled to the APEX telescope. \cite{fasano_spie} }
\label{tab:specs}
\begin{center}       
\begin{tabular}{ll} 
\hline
\rule[-1ex]{0pt}{3.5ex}  Illuminated telescope primary mirror diameter & 11\,m  \\
\hline
\rule[-1ex]{0pt}{3.5ex}  Field-of-view diameter & 18.6\,arcmin   \\
\hline
\rule[-1ex]{0pt}{3.5ex}  Photometric beam widths HF $\mid$ LF & $\sim$30 $\mid$ $\sim$35\,arcsec \\
\hline
\rule[-1ex]{0pt}{3.5ex}  Absolute spectral resolution & settable and down to $\sim$1.5\,GHz \\
\hline
\rule[-1ex]{0pt}{3.5ex}  Frequency range HF $\mid$ LF & 195--310 $\mid$ 130--270\,GHz  \\
\hline
\rule[-1ex]{0pt}{3.5ex}  Pixels on sky HF $\mid$ LF & 2\,152 $\mid$ 2\,152  \\
\hline
\rule[-1ex]{0pt}{3.5ex}  Instrument geometrical throughput & 2.5$\times$10$^{-3}$\,sr$\,$m$^2$  \\
\hline
\rule[-1ex]{0pt}{3.5ex} Single pixel geometrical throughput & 1.16$\times$10$^{-6}$\,sr$\,$m$^2$   \\
\hline
\rule[-1ex]{0pt}{3.5ex}  Data rate & 128\,MB/s  \\
\hline 
\end{tabular}
\end{center}
\end{table}

\begin{figure} [ht]
   \begin{center}
   \begin{tabular}{c} %% tabular useful for creating an array of images 
   \includegraphics[height=6cm]{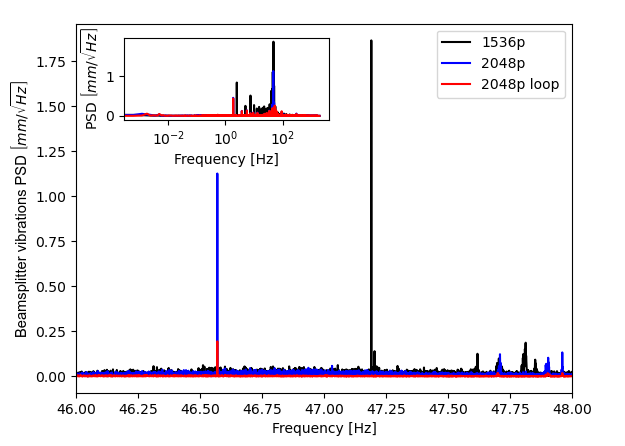}
   \end{tabular}
   \end{center}
   \caption 
   {Power spectral density (PSD) of the beamsplitter oscillation vs frequency. In black is the 1536-point sampling (2.5\,Hz); in blue is the 2048-point one (1.9\,Hz); in red is the 2048-point configuration with the counter system active. The main figure displays the zoom on the frequency range of interest (46--48\,Hz), while the subfigure on the top left shows the whole frequency-domain range (up to 1906.5\,Hz). \cite{fasano_spie}}
   \label{fig:laser3} 
\end{figure}

The complexity of such an instrument demands a precise characterization of its subsystems, a situation which presents peculiar challenges to the interpretation of systematic effects, which, for the most part, have never been tackled before owing to the pioneering nature of the project. These problems must be understood, and innovative solutions have to be found. In CONCERTO, one particular effect concerns the instrument: the roof mirror motion that introduces the OPD in the FTS generates both a wind flow and a vibrational resonance in the central beamsplitter membrane. This effect translates into a complex perturbation of the OPD as a consequence of the large (480$\times$800\,mm$^2$) beamsplitter area, an elliptic 50\,$\mu$m-thick membrane of polyimide on which lie 50\,$\mu$m of copper wires with a pitch of 100\,$\mu$m. This systematic was observed soon after the first commissioning campaign and represented the main issue to be tackled during CONCERTO's commissioning. We needed a solution in order to monitor and mitigate this effect. In December 2021, we installed a laser sensor pointing to the center of the membrane to record the real-time OPD variations. In April 2022, we improved the system with two active speakers programmed to generate counter waves that cancel the vibrations. At the beginning of the commissioning, we adopted the fastest acquisition rate utilized so far by recording a full (forward/backward) motor stroke at 2.5\,Hz. Soon after identifying the membrane vibration issue, we slowed the motion to 1.9\,Hz to reduce the impact of vibration intensity and better sample the interferogram. Finally, we used the slowest motion together with the activation of the counter system. The results of these three set-ups are shown in Fig.~\ref{fig:laser3}, where the vibration measurement is expressed in terms of power spectral density (PSD). The highest speed results in a higher vibration frequency; in contrast, slowing down the motor stroke motion drags the oscillation to a lower frequency with a reduced amplitude. Finally, we successfully managed to mitigate the systematic effect by exploiting a lower speed and the counter system.

\section{Spectral maps of Orion}
\label{sec:orion}

CONCERTO was opened to the scientific community to observe the millimeter Universe, together with the two main scientific drivers. In particular, we present here an observation of the Orion Nebula in spectral mode. This project is part of CONCERTO open time, program number 110.23NK.001, in the ESO period P110. Orion is a reference region for extragalactic astronomy. It epitomizes star-forming regions and is a template for starburst galaxies. The observing program aims to measure dust emissivity in the mm domain \cite{galliano}. The central region of Orion, including the Trapezium \cite{McCaughrean}, has an extent of a quarter of a square degree. The NIKA2 map of the region shows a large extended part of the map above a brightness level of 37\,MJy/sr at 250\,GHz \cite{Ajeddig}. With two hours of observation on-field, a 1-$\sigma$ sensitivity of 0.2 in the emissivity index per 30-arcsecond beam (CONCERTO-like) was obtained.
 
\begin{figure}[h]
\centering
\includegraphics[scale=0.18]{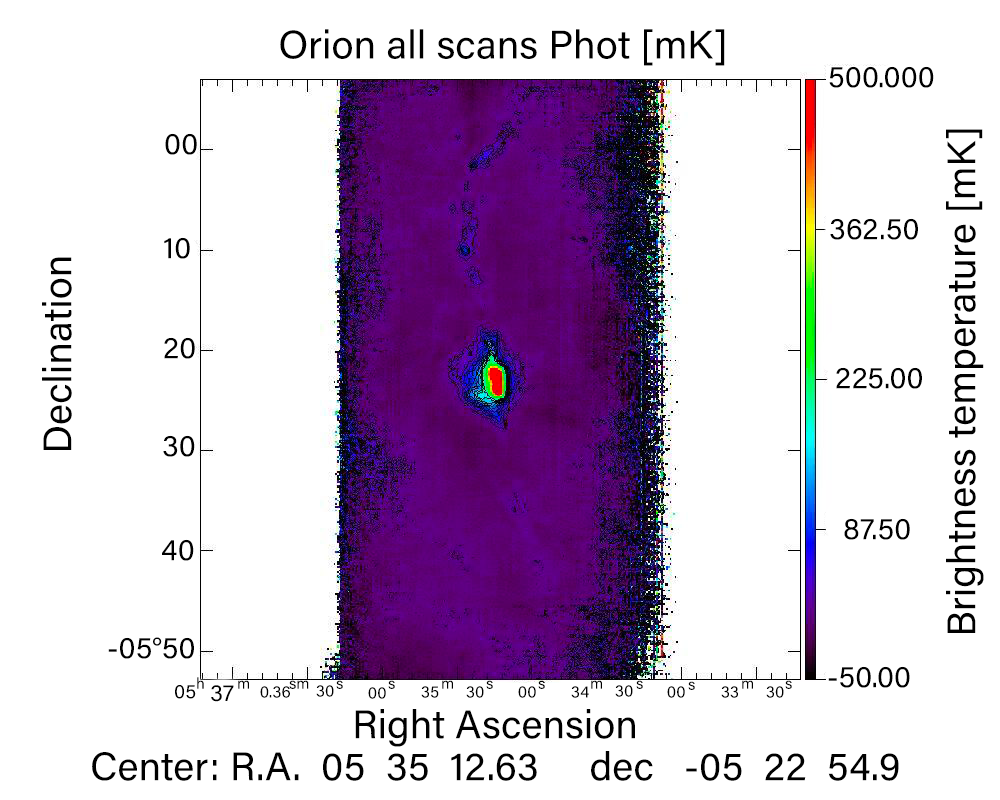}
\includegraphics[scale=0.18]{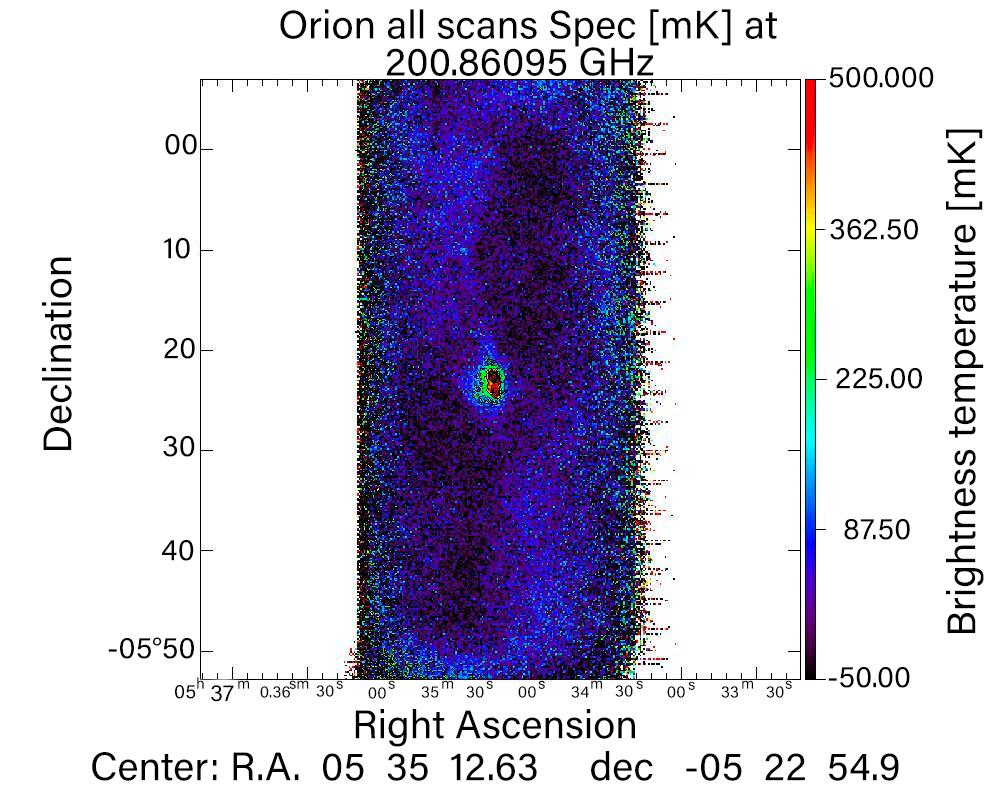}\\
\includegraphics[scale=0.18]{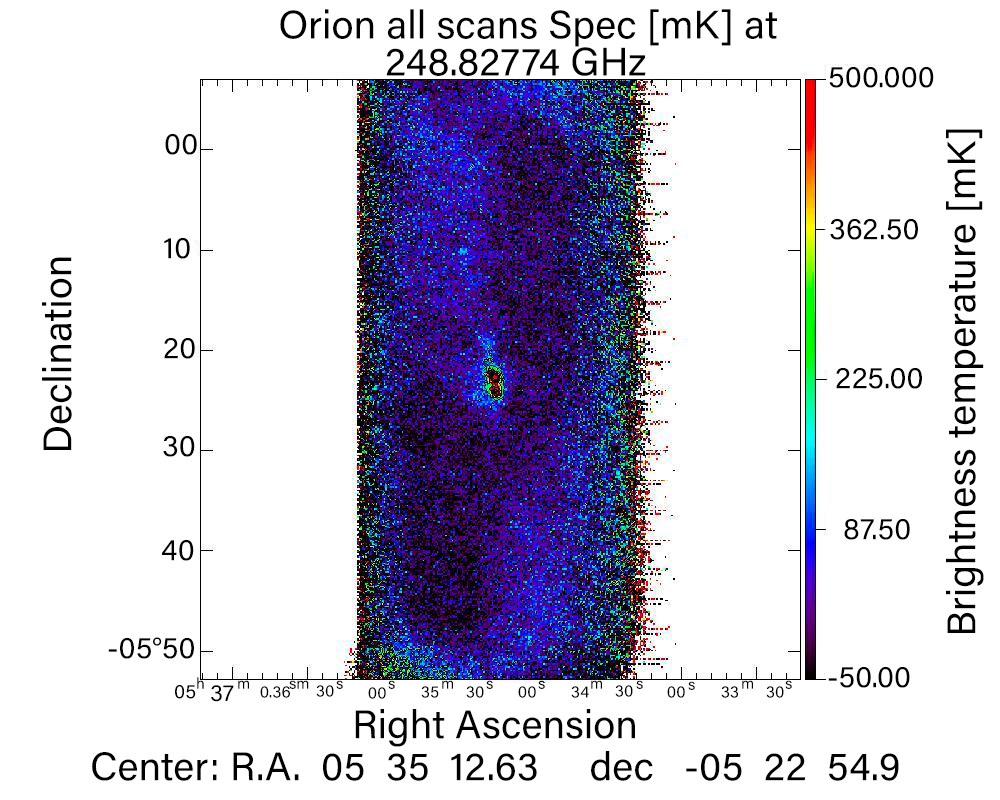}
\includegraphics[scale=0.18]{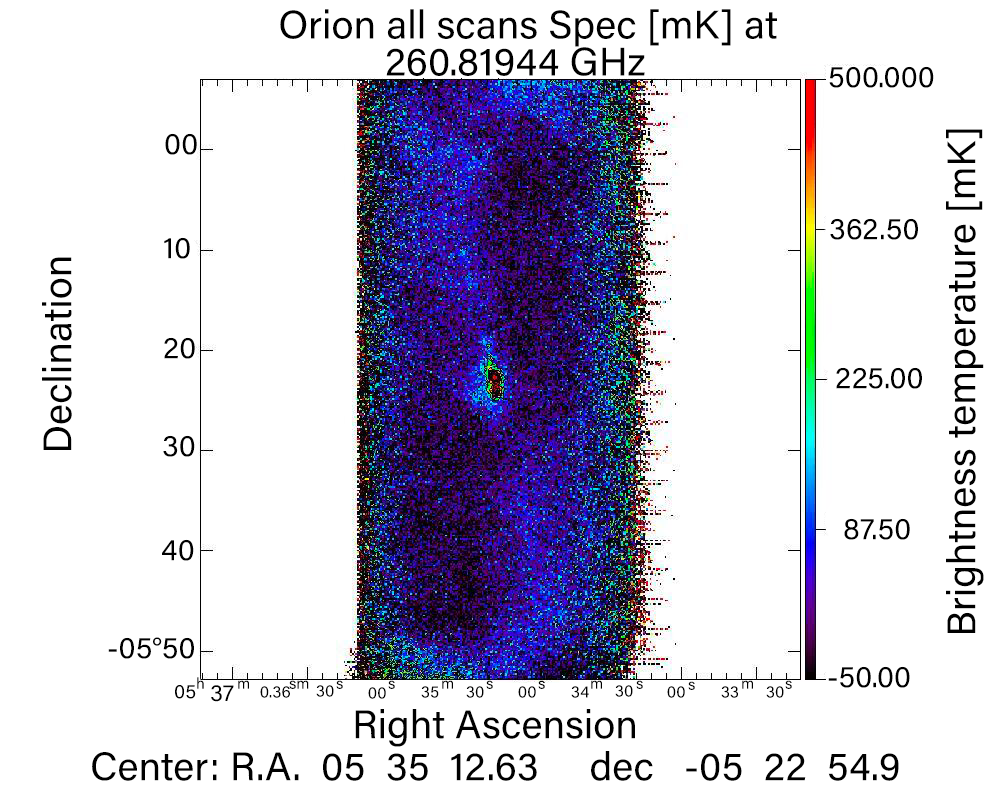}
\caption{Spectral maps of Orion 30$\times$60\,arcmin$^2$. From left to right and top to bottom: 
photometric map of Orion. Bins of 6\,GHz around 200, 246, and 260\, GHz central frequencies. 
These results demonstrate CONCERTO's capability of producing spectral maps of extended regions.}
\label{fig:orion}       
\end{figure}

The Orion Nebula was observed for 2 hours with 12 scans covering an area of 30$\times$60\,arcmin$^2$. The same configuration as for galaxy cluster observations was adopted by setting the CONCERTO FTS stroke at 30\,mm. The data were calibrated using Uranus emission and bandpass measurements. The calibrated photometric map is shown in Fig.~\ref{fig:orion}. At this stage, this preliminary calibration was cross-checked with Planck 217\,GHz map. The CONCERTO map reveals the center of Orion BN/KL, Orion South, the Orion Bar, and the North-South main filament. The Orion Bar is conspicuous in the visible, as demonstrated by Hubble Space Telescope observations. We find that, spectrally, the Bar is dominated by free-free emission, while other regions are dominated by Galactic dust emission. The spectral cube enables us to display Orion with slices at different frequencies (see Fig.~\ref{fig:orion}), covering 140 to 310\,GHz where we avoid contamination by the water vapor line emission at 180\,GHz. These spectral maps represent a sample of CONCERTO's spectral mapping capabilities of extended regions.

\section{Timeline, status and perspectives}
\label{sec:perspectives}

CONCERTO was granted an ERC advanced grant (grant agreement No78812) starting in January 2019. The program was funded to design, fabricate, install, and observe with the CONCERTO instrument at the APEX telescope \cite{gusten}. The instrument was shipped on 1 March 2021, and the team started the installation on 6 April at the 12-m APEX telescope, located at an altitude of 5\,100\,m on the Llano de Chajnantor in Northern Chile.\footnote{\url{https://www.eso.org/public/teles-instr/apex}} After four days (10 April), the cryogenic cooling down started and reached nominal temperature at the focal plane two days after (12 April). The technical commissioning then started. On 2 May, we performed the first extended-source observation (the Crab Nebula\footnote{\url{https://www.eso.org/public/unitedkingdom/announcements/ann21010}}). The observations started two days later remotely from France \cite{monfardini}. The commissioning phase was performed until the end of June 2021 \cite{catalano}. The instrument started its regular scientific program in July 2021. 

CONCERTO was dismounted from the APEX telescope in May 2023 and is back in the Néel Institut (Grenoble, France), in the laboratory where it was designed, fabricated, and assembled. About two years after the shipment, CONCERTO was used for $\sim$793 hours in the [CII] intensity mapping program and for $\sim$465 hours in 11 programs (6 SZ, 3 interstellar medium, and 2 evolved stars). We stocked 50 days of data (174\,TB with a compression factor equal to 5). The collaboration has entered the phase of data exploitation to achieve the scientific targets. Several aspects are being addressed to refine the data processing (e.g., the OPD fine reconstruction, the FTS reference characterization, the KID off-resonance response, and the atmospheric emission contamination). We are also exploiting our development of CONCERTO instrument model, a fundamental tool to properly handle the various systematic effects in terms of interpretation, comprehension, and development of analysis tools to mitigate their effect. Description of the CONCERTO instrument model will be the topic of a forthcoming paper.

In conclusion, this paper has presented the CONCERTO project and instrument, and the delivery of the first spectral maps. Although preliminary in nature, these observations clearly show CONCERTO's considerable potential in the mm domain. These maps, with finer calibration and further study, will be presented in future publications.

\small
{\bfseries \emph{Acknowledgements.}} 
The KID arrays described in this paper have been produced at the PTA Grenoble microfabrication facility. CONCERTO project and this work have been supported by the LabEx FOCUS ANR-11-LABX-0013, the European Research Council (ERC) under the European Union’s Horizon 2020 research and innovation program (project CONCERTO, grant agreement No 788212), and the Excellence Initiative of Aix-Marseille University-A*Midex, a French “Investissements d’Avenir” program.

\end{document}